\documentclass[useAMS,usenatbib]{mnras}
\usepackage{graphicx}
\usepackage{amssymb}
\usepackage{amsmath}

\title[Optimal CCD readout by DCDS]
  {Optimal CCD readout by digital correlated double sampling}
\author[C. Alessandri et al.]
  {C.~Alessandri,$^{1,2}$
  A.~Abusleme,$^1$ D.~Guzman,$^1$
  I.~Passalacqua,$^1$\newauthor E.~Alvarez-Fontecilla,$^{1,3}$
   M.~Guarini,$^1$
  \\
  $^1$ Pontificia Universidad Catolica de Chile, Department of Electrical Engineering, Santiago, Chile.\\
  $^2$ University of Notre Dame, Department of Electrical Engineering, Indiana, United States. \\
  $^3$ University of California San Diego, Department of Electrical and Computer Engineering, California, United States. }
\date{Released 2002 Xxxxx XX}

\pagerange{\pageref{firstpage}--\pageref{lastpage}} \pubyear{2002}

\def\LaTeX{L\kern-.36em\raise.3ex\hbox{a}\kern-.15em
    T\kern-.1667em\lower.7ex\hbox{E}\kern-.125emX}

\begin{document}

\label{firstpage}

\maketitle

\begin{abstract}

Digital correlated double sampling (DCDS), a readout technique for charge-coupled devices (CCD), is gaining popularity in astronomical applications. By using an oversampling ADC and a digital filter, a DCDS system can achieve a better performance than traditional analogue readout techniques at the expense of a more complex system analysis. Several attempts  to analyse and optimise a DCDS system have been reported, but most of the work presented in the literature has been experimental. Some approximate analytical tools have been presented for independent parameters of the system, but the overall performance and trade-offs have not been yet modelled. Furthermore, there is disagreement among experimental results that cannot be explained by the analytical tools available.

In this work, a theoretical analysis of a generic DCDS readout system is presented, including key aspects such as the signal conditioning stage, the ADC resolution, the sampling frequency and the digital filter implementation. By using a time-domain noise model, the effect of the digital filter is properly modelled as a discrete-time process, thus avoiding the imprecision of continuous-time approximations that have been used so far. 

As a result, an accurate, closed-form expression for the signal-to-noise ratio (SNR) at the output of the readout system is reached. This expression can be easily optimised in order to meet a set of specifications for a given CCD, thus providing a systematic design methodology for an optimal readout system. Simulated results are presented to validate the theory, obtained with both time- and frequency-domain noise generation models for completeness.
\end{abstract}


\section{Introduction}

Charge-coupled devices (CCDs) are widely used for scientific imaging because of their high quantum efficiency, linearity and photon dynamic range. However, the dynamic range of astronomical CCDs is usually limited by the readout noise produced by the on-chip amplifier and the reset noise at the sensing capacitor \citep{White1974,Barbe1975,janesick2001}. A correlated double sampling (CDS) scheme removes the reset noise and attenuates low frequency noise components \citep{White1974,Barbe1975}. White noise components can also be reduced by using a limited-bandwidth preamplifier. However, lowering the bandwidth requires a longer separation between samples due to the signal settling, which increases the pixel time and the low-frequency noise contribution \citep{Kansy1980,Hopkinson1982}.

In the search for a better noise reduction, a differential-averaging scheme was proposed, which was proven to be optimal for white noise components \citep{Hegyi1980}. The usual implementation, known as dual slope integration, comprises analogue switches and an integrator \citep{janesick2001}. By using this technique on a standard CCD, the noise can be lowered at the expense of a reduced frame rate by using longer pixel integration times. However, the readout noise cannot be reduced without bound due to the contribution of low-frequency noise, which imposes a noise floor that limits the performance of CCDs for low-light applications. A comprehensive analysis of analogue readout schemes can be found in \cite{Hopkinson1982}, which provides analytical expressions useful for design.

The development of low-noise readout techniques was inactive for over two decades, until \cite{Gach2003} proposed the digital correlated double sampling (DCDS) scheme. In this scheme, most of the analogue circuitry is replaced by an oversampling ADC and a digital filter. Due to the development of high-speed, high-resolution ADCs, the digital implementation of the differential-averaging has outperformed the traditional dual slope integration. Furthermore, the DCDS scheme allows to implement any arbitrarily-shaped filter instead of a simple averaging filter,  thus increasing the design complexity compared to that of the well-studied analogue techniques. 

Based on a qualitative understanding of noise correlation properties, \cite{Gach2003} experimentally found that, for a particular CCD, a weighted filter performs better than an averaging-filter. However, this result was only optimal for a specific setup and was not supported by an analytical framework. Using a different experimental setup, \cite{Clapp2012} tested similar weighted profiles, but reported a better performance for the averaging filter. Clapp also presented an approximated expression to compute the noise of the DCDS system, although it was derived only for an averaging filter. Therefore, the theory failed to explain the disagreement with \cite{Gach2003}. Afterwards, \cite{Tulloch2013} simulated the performance of several weighted filters and reported a marginal noise reduction over the averaging filter at low pixel rates. A first approach to compute optimal weights analytically was presented by \cite{Alessandri2013}, who analysed the design of the digital filter for noise reduction under ideal settling conditions of the video signal. Other design variables such as the ADC sampling frequency and resolution, and the amplifier bandwidth have been studied independently \citep{Tulloch2013,Smith2013,Stefanov2014}. However, there has been no analysis for the overall performance of a DCDS readout system with arbitrary weighted filters.

In this work, an in-depth theoretical analysis of a generic DCDS readout system is presented as follows: Section~\ref{sec:system} provides a mathematical description of the DCDS system. In Section~\ref{sec:stats}, the output statistics of the system are computed with the proper continuous- and discrete-time treatment of the noise processes involved. The SNR optimisation model is depicted in Section~\ref{sec:opt}, and a simulation model for a DCDS readout system is depicted in Section~\ref{sec:sim}. Theoretical and simulated results are presented in Section~\ref{sec:results}. In Section~\ref{sec:conclusion}, conclusions are drawn.

\section{Readout system}

\label{sec:system}
\begin{figure}
   \centering
   \includegraphics[width=\columnwidth]{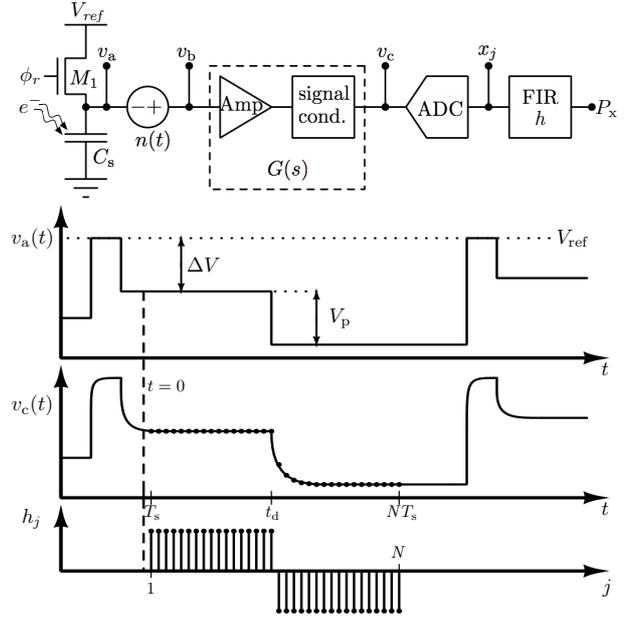} 
   \caption{Generic setup of a DCDS readout system (top), and typical waveforms of a CCD (bottom), where $v_\mathrm{a}$ is the voltage at the CCD sensing capacitor, and $v_\mathrm{c}$ is the voltage after the on-chip amplifier and the signal conditioning stage, both described by $G(s)$. The signal is sampled starting at t=0, and the digital filter depicted by $h_\mathit{j}$ is applied to compute the pixel value. The amplifier noise, modelled by $n(t)$, is not considered in the plots for simplicity.}
   \label{fig:system}
\end{figure}

Fig.~\ref{fig:system} depicts a generic setup of a DCDS readout system along with the characteristic waveforms of a CCD. The measurement of each pixel is performed as follows: the sensing capacitor $C_\mathrm{s}$ is reset to $V_\mathrm{ref}$ by the analogue switch $M_1$. Due to thermal noise, charge injection and clock feedtrough, a voltage drop $\Delta V$ produces an uncertain initial voltage, which will be referred to as the reset voltage. At $t=t_\mathrm{d}$, the pixel charge is transferred to the sensing capacitor, discharging the capacitor by a voltage $V_\mathrm{p}$, which is related to the pixel charge $n_\mathrm{e}$ by the output sensitivity $S_\mathrm{v}$, thus
\begin{eqnarray}
V_\mathrm{p}=S_\mathrm{v}n_\mathrm{e}.
\label{eq:vs}
\end{eqnarray}
 Therefore, the voltage at the sensing capacitor can be expressed as
\begin{eqnarray}
v_\mathrm{a}(t)=V_\mathrm{r}-v_\mathrm{p}(t),
\label{eq:vs}
\end{eqnarray}
where \mbox{$V_\mathrm{r}=V_\mathrm{ref}-\Delta V$} is the reset voltage, \mbox{$v_\mathrm{p}(t)=V_\mathrm{p}u(t-t_\mathrm{d})$} is the pixel signal and $u(t)$ is the Heaviside function. The reset pulse is left out of equation~\eqref{eq:vs} for simplicity, and it is assumed that the reset voltage is fully settled. 

The voltage at the sensing capacitor is buffered by an on-chip amplifier, which adds noise to the measurement. This amplifier can be modelled as a noiseless amplifier (block Amp in Fig.~\ref{fig:system}) preceded by an equivalent series noise voltage  source with two-sided PSD $S(i\omega)$ \citep{gray2009}. Hence, the voltage at the input of the noiseless amplifier is given by
\begin{eqnarray}
v_\mathrm{b}(t)=V_\mathrm{r}-v_\mathrm{p}(t)+n(t),
\label{eq:vamp}
\end{eqnarray}
where $n(t)$ is the amplifier input-referred series noise voltage.

The CCD output is processed by a signal conditioning stage as depicted in Fig.~\ref{fig:system}. For analysis purposes, the noiseless amplifier and the signal conditioning circuit can be described by a single generic transfer function $G(s)$ with impulse response $g(t)$.

The signal at the ADC input can be computed as a linear convolution between $v_\mathrm{b}(t)$ and $g(t)$, hence
\begin{eqnarray}
v_\mathrm{c}(t)		&=& \{V_\mathrm{r}*g\}(t)-\{v_\mathrm{p}*g\}(t)+\{n*g\}(t).
\label{eq:vadc}
\end{eqnarray}
Then, the signal is sampled with a period $T_\mathrm{s}$, where $t=0$ is arbitrarily defined before the first sample, as shown in Fig.~\ref{fig:system}. A column vector of $\mathit{N}$ samples $\textbf{\textit{x}}=[x_1,.., x_\mathit{j},..,x_\mathit{N}]^t$ is taken at $t=jT_\mathrm{s}$, with $j=1..\mathit{N}$, thus 
\begin{eqnarray}
x_\mathit{j} &=& v_\mathit{j}+r_\mathit{j}+n_\mathit{j}+q_\mathit{j} \label{eq:xk},
\end{eqnarray}
where \mbox{$\!v_\mathit{j}\!=\! -\{v_\mathrm{p}\!*\!g\}(jT_\mathrm{s})\!$}, \mbox{$\!r_\mathit{j}\!=\! \{V_\mathrm{r}\!*\!g\}(jT_\mathrm{s})\!$}, \mbox{$\!n_\mathit{j}\!=\!\{n\!*\!g\}(jT_\mathrm{s})\!$} and $q_\mathit{j}$ is the quantisation and electronic noise introduced by the ADC. Finally, the digital filter described by \mbox{$\textbf{\textit{h}}\!=\![h_1,.., h_\mathit{j},..,h_\mathit{N}]^t$} is applied to compute the pixel value as 
\begin{eqnarray}
P_\mathrm{x} &=& \textbf{\textit{h}}^t\textbf{\textit{x}} \nonumber \\
       &=& \sum_{\mathit{j}=1}^\mathit{N} h_\mathit{j}x_\mathit{j}, 
\end{eqnarray}
which is the output of the DCDS system. 

\section{Output statistics}
\label{sec:stats}

Knowing the noise characteristics of the system at the ADC input, the expression for the SNR after the digital filter is derived as follows. The mean value of the pixel measurement can be computed as
\begin{eqnarray}
           \mu_\mathrm{x}        &=& \mathrm{E}\{P_\mathrm{x}\} \nonumber\\
			   &=& \sum_{\mathit{j}=1}^\mathit{N}h_\mathit{j}(\mathrm{E}\{v_\mathit{j}\}+\mathrm{E}\{r_\mathit{j}\}+\mathrm{E}\{n_\mathit{j}\}+\mathrm{E}\{q_\mathit{j}\}) \nonumber\\
			   &=& \sum_{\mathit{j}=1}^\mathit{N}h_\mathit{j}(v_\mathit{j}+r_\mathit{j}) \label{eq:mean},
\end{eqnarray}
since $n_\mathit{j}$ and $q_\mathit{j}$ are zero-mean random variables (see Section \ref{sec:adcnoise}), and both $v_\mathit{j}$ and $r_\mathit{j}$ are deterministic functions of $V_\mathrm{r}$ and $V_\mathrm{p}$, which are constant within a pixel. The variance of the pixel measurement is given by
\begin{eqnarray}
						\sigma_\mathrm{x}^2	&=& \mathrm{E}\{\left(\textbf{\textit{h}}^t\textbf{\textit{x}}-\mu_\mathrm{x}\right)^2\} \nonumber \\
									&=& \mathrm{E}\left\{\left(\sum_{\mathit{j}=1}^\mathit{N}h_\mathit{j}n_\mathit{j}+h_\mathit{j}q_\mathit{j}\right)^2\right\}. 
\end{eqnarray}
Considering that $n_\mathit{j}$ and $q_\mathit{j}$ are independent variables (see section \ref{sec:adcnoise}), the expected value of their product is zero, thus 

\begin{eqnarray}
						\sigma_\mathrm{x}^2	
									&=& \sum_{\mathit{j}=1}^{N}\sum_{\mathit{k}=1}^{N}h_\mathit{j}h_\mathit{k}\mathrm{E}\{n_\mathit{j}n_\mathit{k}\}+\sum_{\mathit{j}=1}^{N}\sum_{\mathit{k}=1}^{N}h_\mathit{j}h_\mathit{k}\mathrm{E}\{q_\mathit{j}q_\mathit{k}\} \nonumber \\
									&=& \sum_{\mathit{j}=1}^{N}\sum_{\mathit{k}=1}^{N}h_\mathit{j}h_\mathit{k}\mathrm{R}_\mathrm{n}[j,k]+\sum_{\mathit{j}=1}^{N}\sum_{\mathit{k}=1}^{N}h_\mathit{j}h_\mathit{k}\mathrm{R}_\mathrm{q}[j,k]  \nonumber \\
									&=& \sigma_\mathrm{amp}^2+\sigma_\mathrm{ADC}^2, \label{eq:auc}
\end{eqnarray}
where $\mathrm{R}_\mathrm{n}[j,k]$ and $\mathrm{R}_\mathrm{q}[j,k]$ are the terms of the discrete autocorrelation matrices of the amplifier and ADC noise, respectively. The noise models for these processes and the procedures to compute $\sigma_\mathrm{amp}^2$ and $\sigma_\mathrm{ADC}^2$  are presented separately in the following subsections.

\subsection{Output amplifier noise}

The noise of the CCD output amplifier usually comprises white noise and one or more low-frequency noise components \citep{janesick2001, Hopkinson1982}. For mathematical purposes, the two-sided PSD of the amplifier input-referred series noise voltage is described as a superposition of power-law noise sources given by
\begin{eqnarray}
S(i\omega)		&=&\sum_\mathrm{m}A_\mathrm{m}\left\vert \omega \right\vert^{\alpha_\mathrm{m}}      [\mathrm{V}^2/\mathrm{Hz}] \nonumber \\
				&=&\sum_\mathrm{m}S_\mathrm{m}(i\omega)   [\mathrm{V}^2/\mathrm{Hz}] ,
\label{eq:PSD2}
\end{eqnarray} 
which describes white noise ($\alpha_\mathrm{m}=0$) and low-frequency noise, where $\alpha_\mathrm{m}$ is usually between -1 and -2.  Accordingly, at the ADC input, the noise spectrum is given by
\begin{eqnarray}
S_\mathrm{c}(i\omega)=\sum_\mathrm{m}S_\mathrm{m}\left(i \omega \right)\left\vert G(i\omega)\right\vert  ^2    [\mathrm{V}^2/\mathrm{Hz}] .
\label{eq:PSD2b}
\end{eqnarray} 

Given the composition of equation \eqref{eq:PSD2b}, the output-referred voltage noise will be derived for a single power-law noise source $S_\mathrm{m}(i\omega)$, and the total noise can be computed as the superposition in quadrature of the contribution of each power-law noise source.

Although the autocorrelation matrix from equation \eqref{eq:auc} could be computed by the inverse Fourier transform of $S_\mathrm{c}(i\omega)$, it usually does not yield a closed-form expression and requires $\mathit{N}$ infinite-length numerical integrations. Therefore, the resulting expression for $\sigma_\mathrm{x}^2$ provides little insight for design. An alternative approach, widely used in instrumentation for detectors in particle physics experiments, employs a time-domain noise model to design optimal filters. The noise is modelled as a sequence of pulses with a certain shape $\tilde{y}(t)$, arriving poissonianly at times $t_\mathrm{a}$ with an average rate $\nu$ and random sign \citep{Avila2013,goulding1972, Radeka1988, pullia2001,pullia2004}. The pulse shape that models a noise source $S_\mathrm{m}\left(i \omega \right)$ referred to the ADC input is expressed as (see appendix \ref{ap:pulse})
\begin{eqnarray}
\tilde{y}_\mathrm{m}(t)=\sqrt{\frac{A_\mathrm{m}}{\nu }}\frac{\mathrm{d}^{\alpha_\mathrm{m}/2}}{\mathrm{d}t^{\alpha_\mathrm{m}/2}} g(t).
\label{eq:coretimea}
\end{eqnarray}

The total integrated noise $\sigma^2_\mathrm{m}$ measured at the ADC input is computed in the time domain using Campbell theorem \citep{papoulis2002}. 
\begin{eqnarray}
\sigma^2_\mathrm{m}&=&\nu\int_{-\infty}^{t}{\tilde{y}_\mathrm{m}^2}(t-t_\mathrm{a})dt_\mathrm{a} \nonumber \\
			       &=&\int_{-\infty}^{\infty}{{y}_\mathrm{m}^2}(t_\mathrm{a})dt_\mathrm{a}, 
\label{eq:sigmatb}
\end{eqnarray}
which is equivalent to the amplifier noise autocorrelation function evaluated at $t=0$ (see appendix \ref{ap:auto}). When the noise converges to a finite value, and according to Parseval theorem, $\sigma^2_\mathrm{m}$ can also be computed in the frequency domain \citep{Radeka1988}, thus
\begin{eqnarray}
\sigma^2_\mathrm{m}&=&\frac{1}{2\pi} \int_{-\infty}^{\infty}{S_\mathrm{m}}(i\omega)\left\vert G(i\omega)\right\vert^2d\omega.
\label{eq:sigmat2}
\end{eqnarray}
The total integrated noise can be decomposed into two uncorrelated noise sources: the noise contribution of pulses that arrive before sampling (i.e. $t_\mathrm{a}<0$) and the noise generated within the sampling window (i.e. $0<t_\mathrm{a}<\mathit{N}T_\mathrm{s}$), hence
\begin{eqnarray}
\sigma^2_\mathrm{m} &=& \int_{-\infty}^{0}{y_\mathrm{m}^2}(t-t_\mathrm{a})dt_\mathrm{a} + \int_{0}^{t}{y_\mathrm{m}^2}(t-t_\mathrm{a})dt_\mathrm{a} \nonumber \\
				&=& \sigma_{\mathrm{m},0}^2(t)			+ \sigma_{\mathrm{m},t}^2(t) \label{eq:sigmat2}.
\label{eq:sigmas}
\end{eqnarray}
Since $\sigma_{\mathrm{m},0}^2(t)$ is the contribution of pulses generated before the first sample, its autocorrelation matrix is given by
\begin{eqnarray}
\mathrm{R}_{\mathrm{m},0}[j,k] = \sigma_{\mathrm{m},0}(jT_\mathrm{s})\sigma_{\mathrm{m},0}(kT_\mathrm{s}),
\label{eq:stationary}
\end{eqnarray}
and its contribution after the filter is directly computed as
\begin{eqnarray}
\hat{\sigma}_{\mathrm{m},0}^2 &=& \left(\sum_{\mathit{j}=1}^{N} h_\mathit{j}\sigma_{\mathrm{m},0}(jT_\mathrm{s})\right)^2. 
\label{eq:stationary}
\end{eqnarray}
Using equation~\eqref{eq:sigmas},  this can be written as
\begin{eqnarray}
\hat{\sigma}_{\mathrm{m},0}^2 &=& \left(\sum_{\mathit{j}=1}^{N} h_\mathit{j}\sqrt{\sigma_\mathrm{m}^2-\sigma_{\mathrm{m},t}^2(jT_\mathrm{s})}\right)^2.
\label{eq:stationary}
\end{eqnarray}

The contribution of the noise generated within the sampling window is computed by the same principle, which is developed in detail by \cite{Avila2013}. Thus,
\begin{eqnarray}
\hat{\sigma}_{\mathrm{m},t}^2= \sum_{\mathit{j}=1}^\mathit{N} \left( \sum_{\mathit{k}=0}^{\mathit{N}-\mathit{j}} h_\mathit{\mathit{j}+\mathit{k}}\sqrt{\sigma_{\mathrm{m},t}^2((k+1)T_\mathrm{s})-\sigma_{\mathrm{m},t}^2(kT_\mathrm{s})}\right)^2\hspace{-0.5em}. \label{eq:final}
\end{eqnarray}
Finally, the output referred contribution of $S_\mathrm{m}(i\omega)$ is 
\begin{eqnarray}
\hat{\sigma}_\mathrm{m}^2 = \hat{\sigma}_{\mathrm{m},0}^2 +\hat{\sigma}_{\mathrm{m},t}^2
\end{eqnarray}
 and the total amplifier noise contribution is added in quadrature, hence
\begin{eqnarray}
{\sigma}_\mathrm{amp}^2 =  \sum_\mathrm{m} \hat{\sigma}_\mathrm{m}^2.
\label{eq:ampnoise}
\end{eqnarray}

\subsection{ADC noise autocorrelation}
\label{sec:adcnoise}
Consider an ADC with a resolution of $B$ bits and a full-scale voltage range $V_\mathrm{FSR}$, so $\Delta=V_\mathrm{FSR}/2^B$ is the least-significant bit (LSB). If the ADC is not overloaded, and if the input signal is large and active enough to span over several quantisation levels,  the quantisation noise is modelled as an uncorrelated, zero-mean white noise with variance $\sigma_q^2={\Delta}^2/12$ \citep{Widrow1956}. In the case of a DCDS system, a slow-varying but noisy signal is sampled, and the aforementioned conditions are met if $\Delta$ is comparable to the standard deviation of the independent noise between two samples. This noise is composed by the CCD noise contribution generated within two samples and the ADC electronic noise $\sigma_e^2$, also called transition noise. Therefore, the LSB is upper-limited by
\begin{eqnarray}
\Delta<\sqrt{\sigma_{e}^2 + \sum_\mathrm{m}\sigma^2_{\mathrm{m},t}(T_\mathrm{s})}.
\label{eq:ADCcond}
\end{eqnarray}
Under this assumption, the autocorrelation matrix of the ADC noise is given by
\begin{eqnarray}
\mathrm{R}_\mathrm{q}[j,k]=\delta[j,k]\left(\sigma_\mathrm{q}^2+\sigma_\mathrm{e}^2\right),
\label{eq:adcnoise}
\end{eqnarray}
where $\delta[j,k]$ is the Kronecker delta. The ADC noise contribution at the filter output is directly computed  as
\begin{eqnarray}
{\sigma}^2_\mathrm{ADC}=\left(\sigma_\mathrm{q}^2+\sigma_\mathrm{e}^2\right)\sum_{\mathit{j}=1}^\mathit{N}h_\mathit{j}^2.
\label{eq:adcnoise}
\end{eqnarray}
For larger values of $\Delta$, the quantisation noise may be partially correlated and the noise contribution will be higher than that predicted in equation~\eqref{eq:adcnoise}. Therefore, in order to benefit from the quantisation noise reduction of the digital filter, the ADC resolution is lower-limited by 
\begin{eqnarray}
B>\mathrm{log}_2\left(\frac{V_\mathrm{FSR}}{\sqrt{\sigma_{e}^2 + \sum_\mathrm{m}\sigma^2_{\mathrm{m},t}(T_\mathrm{s})}}\right).
\label{eq:ADCcond2}
\end{eqnarray}
Nevertheless, a higher resolution still provides a benefit in the optimal setup due to a lower quantisation noise, and equation~\eqref{eq:ADCcond2} is rarely an active restriction in low-noise applications. Furthermore, typical high-resolution ADCs have a transition noise of several LSB, so this equation is met regardless of the CCD noise. If the ADC resolution is fixed, the full-scale range referred to the sensing capacitor can be adjusted by the gain at the signal conditioning stage, thus trading the electrons range for a lower quantisation noise.
Although there are more thorough models for the quantisation noise autocorrelation matrix \citep{Gray1990,Gray1998}, the model presented here is accurate for the conditions of operation of a DCDS system and was chosen for its simplicity.

\section{SNR optimisation}
\label{sec:opt}
In order to optimise the SNR, an analytical expression for the impulse response of the signal conditioning stage should be given, since it determines both the mean value and the variance of the pixel measurement. A typical signal conditioning stage for a DCDS system has a transfer function of the form
\begin{eqnarray}
G(s) =G_0 \frac{\tau_2s}{(1+\tau_2s)(1+\tau_1s)}    ,
\label{eq:G(S)} 
\end{eqnarray}
which comprises a single-pole high-pass filter defined by $\tau_2$, static gain $G_0$ and a single-pole low-pass filter with time constant $\tau_1$. However, it is straightforward to extend the analysis presented here for higher-order systems. 

Even though $G(s)$ comprises the effect of the AC coupling capacitor, in a well-designed system the coupling capacitor will be large enough so as to keep the signal integrity within a pixel \citep{Hegyi1980}. Hence \mbox{$G(s)\approx G_0 /\left(1+\tau_1s \right)$}. By setting $t_\mathrm{d}=\frac{N}{2}T_\mathrm{s}$, and 
according to equation~\eqref{eq:mean}, the pixel mean value is
\begin{eqnarray}
\mu_\mathrm{x}\!=\!G_0\!\!\left(\!\!V_\mathrm{p}\!\!\!\!\sum_{\mathit{j}=\frac{N}{2}+1}^{N}{\!\!\!\!h_\mathit{j}\!\!\left(1\!-\!e^{-{\left(j-\frac{N}{2}\right)T_\mathrm{s}/\tau_1}}\right)}+V_\mathrm{r}\sum_{\mathit{j}=1}^{N}{h_\mathit{j}}\right).
\label{eq:signal}
\end{eqnarray}
Since the reset voltage remains constant within a pixel, it can be completely removed if the filter coefficients add up to zero, which is the basis of the differential sampling scheme. Replacing the signal conditioning impulse response into equation \eqref{eq:coretimea}, and computing the fractional derivative, the pulse shape ${{y}}_\mathrm{m}(t)$ can be expressed as
\begin{eqnarray}
{y}_\mathrm{m}(t)\!=\!\sqrt{A_\mathrm{m}}G_0u(t)\!\!\!\!\!\!\!\!&\left(\frac{\tau_2/\tau_1}{\tau_2-\tau_1}t^{-\alpha_\mathrm{m}/2}E_{1,1-\alpha_\mathrm{m}/2}(-t/\tau_1) \right.  \nonumber\\
		      &\left.\hspace{3pt}-\frac{1}{\tau_2-\tau_1}t^{-\alpha_\mathrm{m}/2}E_{1,1-\alpha_\mathrm{m}/2}(-t/\tau_2)\right),
\label{eq:coretimeeval}
\end{eqnarray}
where $E_{a,b}(t)$ is the Mittag-Leffler function \citep{mathai2008}. 
Finally the SNR is expressed as
\begin{eqnarray}
\mathrm{SNR}=\frac{\left(G_0V_\mathrm{p}\sum_{\mathit{j}=\frac{N}{2}+1}^{N}{h_\mathit{j}\left(1-e^{-{\left(j-\frac{N}{2}\right)T_\mathrm{s}/\tau_1}}\right)}\right)^2}{\sigma^2_\mathrm{amp}+\sigma^2_\mathrm{ADC}},
\label{eq:SNR}
\end{eqnarray}
which is an analytic function of the CCD noise parameters, the filter coefficients and a set of design variables \mbox{${\boldsymbol\gamma}=\{G_0, \tau_1,\tau_2, T_\mathrm{s}, N, B, V_\mathrm{FSR}\}$}. The signal power, the reset noise and the amplifier noise are proportional to $G_0^2$, therefore changing the gain only affects the overall SNR due to the quantisation noise.

The optimisation is performed as follows. Given a fixed set of design variables \mbox{$\tilde{\boldsymbol\gamma}=\{\tilde{G_0}, \tilde{\tau_1}, \tilde{T_\mathrm{s}}, \tilde{N}, \tilde{B}, \tilde{V}_\mathrm{FSR}\}$}, the noise coefficients $\sigma_\mathrm{m}^2$ and $\sigma_{\mathrm{m},t}^2(jT_\mathrm{s})$ can be pre-computed with a single, finite-length numerical integration, and the SNR  can be expressed solely as a function of the filter coefficients. Since the SNR is a highly nonlinear function, the filter optimisation is carried out by fixing the pixel gain and minimizing the noise. Hence, the optimisation problem is formulated as:

\begin{eqnarray}
\begin{aligned}
& \underset{\textbf{\textit{h}}}{\text{minimise}}
& & \sigma^2_\mathit{read}(\textbf{\textit{h}},\tilde{\boldsymbol\gamma}) =  \sigma^2_\mathrm{amp}+\sigma^2_\mathrm{ADC}\\
& \text{subject to} & \\
& & & \!\!\!\!\!\sum_{\mathit{j}=\frac{N}{2}+1}^{N} {\!\!\!\!h_\mathit{j}\left(1-e^{-{\left(j-\frac{N}{2}\right)\tilde{T_\mathrm{s}}}/{\tilde{\tau_1}}}\right)}={1}{} \\
& & & \sum_{\mathit{j}=1}^{N}{h_\mathit{j}}=0.
\end{aligned}
\end{eqnarray}
This problem can be solved with standard optimisation software tools \citep{AMPL,Knitro}. The overall optimisation is performed as a semi-exhaustive search in the design space ${\boldsymbol\gamma}$, which is usually bounded by the application requirements, available hardware and other design-related trade-offs.

\section{DCDS readout system simulation setup}
\label{sec:sim}

\begin{figure}
   \centering
   \includegraphics[width=\columnwidth]{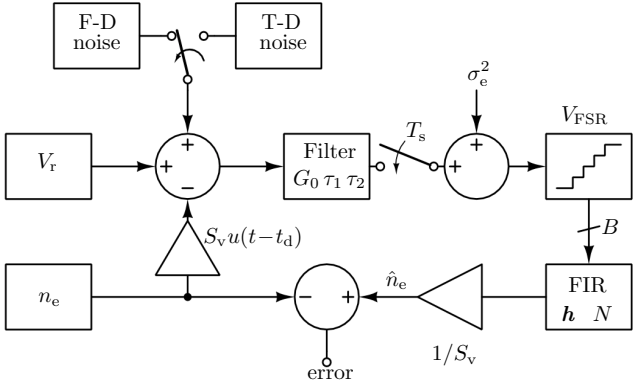} 
   \caption{Simulation diagram: for each pixel, the reset voltage, pixel charge and amplifier noise are randomly generated and added as voltages. Time- and frequency-domain models can be selected for noise generation. An analogue filter is emulated with the simulation time-step, and then the signal is down-sampled to $T_\mathrm{s}$. The ADC electronic noise is added and the signal is quantised. Finally, the digital filter is applied to compute the error.}
   \label{fig:simulation}
\end{figure}

Based on the mathematical description of the DCDS readout system presented in Section~\ref{sec:system}, a set of simulations were programmed in {\tiny MATLAB}.  As depicted in Fig.~\ref{fig:simulation}, a random reset voltage $V_\mathrm{r}$ is generated for each pixel. The pixel charge is computed as a random, integer number of electrons $n_\mathrm{e}$, which is converted into voltage with the output sensitivity and added to the reset voltage at $t=t_\mathrm{d}$. The amplifier PSD is defined by white noise and a single low-frequency noise component, hence
\begin{eqnarray}
S(i\omega)=A_s+A_f\vert \omega \vert^b,
\end{eqnarray}
with $-2\leq b \leq -1$. It is usual to describe the low-frequency noise amplitude by the corner frequency $f_\mathrm{c}$, defined as the frequency at which the low-frequency noise power is equal to the white noise power. In this case,
\begin{eqnarray}
S(i\omega)=A_s\left(1+\left\vert \frac{\omega}{2\pi f_\mathrm{c}} \right\vert^b\right)
\label{eq:swfc}
\end{eqnarray}
and $A_\mathrm{f}=A_\mathrm{s}(2\pi f_\mathrm{c})^{-b}$.

For completeness, the noise can be generated by two methods:
\begin{itemize}
\item Time-domain (T-D) generation of noise pulses, based on the method proposed by \cite{pullia2004}.
\item Frequency-domain (F-D) generation of noise, implemented by the method proposed by \cite{Kasdin1995}. 
\end{itemize}
The noise is added to the signal, and the analogue filter, described by $G_0,\tau_1$ and $\tau_2$, is emulated to obtain the signal at the ADC input. The time-step of the simulation is defined by an oversampling rate over $T_\mathrm{s}$ for accuracy in the noise generation and filtering, so the signal is downsampled to $T_\mathrm{s}$  at the ADC to generate $N$ samples. The ADC electronic noise is added to these samples, which are quantised with resolution $B$ over a voltage range $V_\mathrm{FSR}$ and digitally filtered by the FIR described by $\textbf{\textit{h}}$. The pixel value is converted to electrons and compared with $n_\mathrm{e}$ to compute the error. 
The simulation is entirely determined by the design variables \mbox{${\boldsymbol\gamma}=\{G_0, \tau_1,\tau_2, T_\mathrm{s}, N, B, V_\mathrm{FSR}\}$}, the filter coefficients and the system parameters \mbox{$\boldsymbol{{\zeta}}=\{A_\mathrm{s},A_\mathrm{f},b,S_\mathrm{v},\sigma_\mathrm{e}\}$}.

\section{Theoretical and simulated results}
\label{sec:results}

\begin{table}
\centering
\begin{tabular}{ccc}
\hline 
Parameter & CCD1 & CCD2 \\ \hline
 $A_\mathrm{s}$	 	& $\left(0.5\frac{nV}{\sqrt{Hz}}\right)^2$ 	& $\left(1.5\frac{nV}{\sqrt{Hz}}\right)^2$ \\
 $f_\mathrm{c}$ 		& $20\mathrm{kHz} $			   	& $150\mathrm{kHz} $ \\
 $b$	 				& -1.2 & -1\\
 $S_\mathrm{v}$ 		& $2.5 \mu V/e^-$ & $8 \mu V/e^-$\\
 \hline
\end{tabular}
\caption{CCD1 and CCD2 noise parameters and sensitivity. The noise PSD is described by equation~\eqref{eq:swfc}.}
\label{tab:params}
\end{table}

\begin{figure}
   \centering
   \includegraphics[width=\columnwidth]{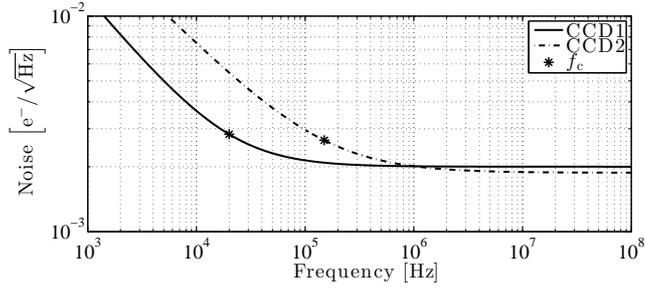} 
   \caption{Noise PSD of CCD1 and CCD2. The noise amplitude is referred to the sensing capacitor by the sensitivity $S_\mathrm{v}$ and shown in units of $\mathrm{e}^-/\sqrt{\mathrm{Hz}}$ for a fair comparison. The low-frequency noise corner frequency $f_\mathrm{c}$ is marked for each CCD.}
   \label{fig:spectrums}
\end{figure}

A set of theoretical and simulated results are presented to validate the theory and illustrate the potential of the proposed method. The results were generated for the two sets of parameters shown in Table~\ref{tab:params}, which are characterised by the noise PSD depicted in Fig.~\ref{fig:spectrums}. The CCD1 parameters were estimated from \cite{Cancelo2012}, whereas the parameters for CCD2 were taken from \cite{Tulloch2013}, which depicts a typical E2V CCD. The LSB is set to 1 electron, so a full-well of up to 262.144 electrons could be read for an 18-bit ADC, and the ADC electronic RMS noise $\sigma_\mathrm{e}$ was set at $3\Delta$. The high-pass filter time constant is fixed at $10$ Hz to keep the signal integrity. 

\begin{figure}
   \centering
   \includegraphics[width=\columnwidth]{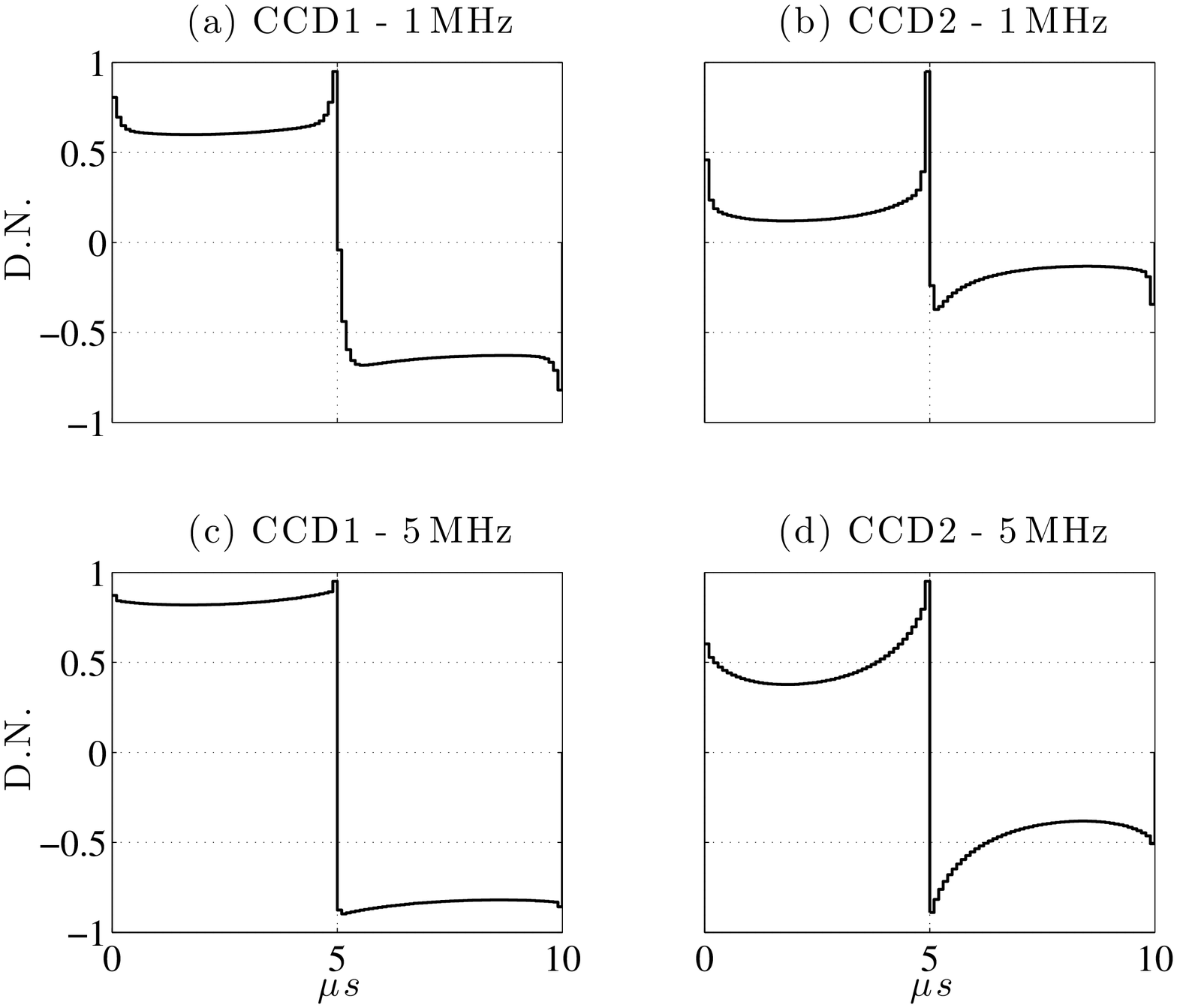} 
   \caption{Normalised filter coefficients for a $10\mathrm{\mu s}$ sampling window and 100 samples. The optimal coefficients were computed for both CCDs with $1$ and $5$ MHz bandwidths. }
   \label{fig:coeffs_fc}
\end{figure}

\begin{figure}
   \centering
   \includegraphics[width=\columnwidth]{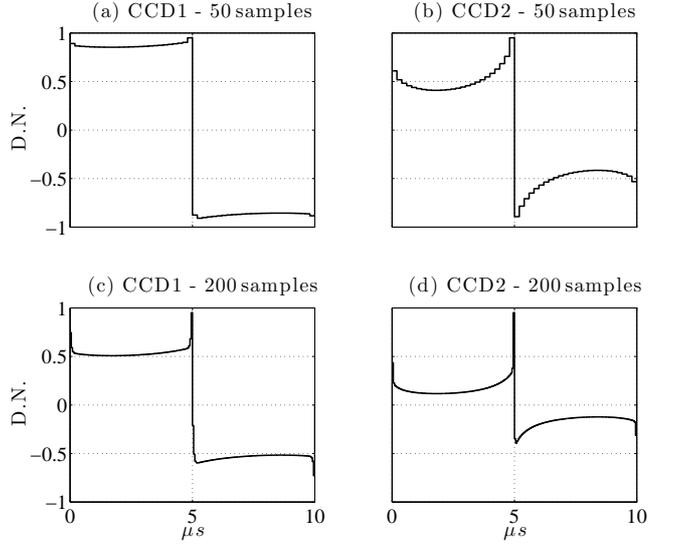} 
   \caption{Normalised filter coefficients for a $10\mathrm{\mu s}$ sampling window and 2.5 MHz bandwidth. The optimal coefficients were computed for both CCDs with 50 and 200 samples. }
   \label{fig:coeffs_Nsamp}
\end{figure}

Figs. \ref{fig:coeffs_fc} and \ref{fig:coeffs_Nsamp} show a set of optimal filter coefficients for different scenarios. Since CCD2 has a higher corner frequency than CCD1, the optimal coefficients for CCD2 are always steeper near the charge dump, which is consistent with the principle introduced by \cite{Gach2003}. Figs.~\ref{fig:coeffs_fc}~(a) and (b) show that the coefficients are not symmetrical for low bandwidths, whereas for a higher bandwidth as in Fig.~\ref{fig:coeffs_fc}~(c) and (d), the optimal filter approaches those already reported in the literature for ideal signal setting \citep{Alessandri2013}. These results can be understood by considering that a lower bandwidth enlarges the noise temporal correlation, thus producing a better noise cancellation by the subtraction near the charge dump. Therefore, the optimal solution assigns more weights to the middle coefficients. However, some samples after the charge dump are attenuated because the charge is not fully settled, thus there is an optimal bandwidth for this trade-off. In this case, for both CCDs the noise performance was better at 1 MHz. This approach defies the accepted convention to use a high bandwidth and discard samples until the signal is settled after the charge dump. Imposing these conditions, the optimal coefficients tend to be flat but produce a sub-optimal result due to the additional restrictions. This explains the disagreement between \cite{Gach2003} and \cite{Clapp2012}, and supports the results reported by \cite{Tulloch2013}.

\begin{figure}
   \centering
   \includegraphics[width=\columnwidth]{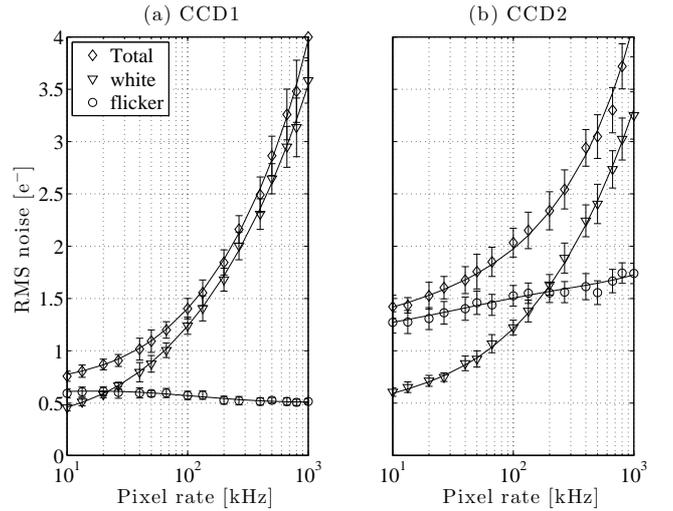} 
   \caption{RMS noise along with white and flicker noise contributions vs pixel rate. The results were generated with a 40 MSPS ADC and 500 kHz bandwidth. The theoretical predictions are plotted with solid lines and the simulation results are shown by the error bars.}
   \label{fig:noises}
\end{figure}

\begin{figure}
   \centering
   \includegraphics[width=\columnwidth]{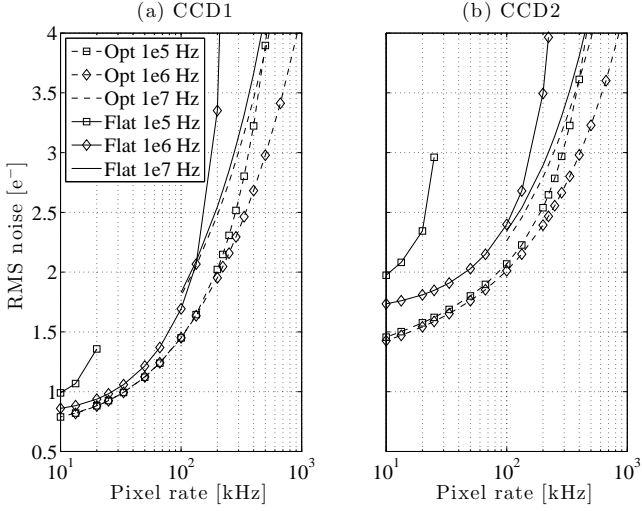} 
   \caption{RMS noise vs pixel rate for both CCDs. The standard averaging filter (Flat) is compared with the optimal filter computed by the proposed method (Opt). The results were generated with a 20 MSPS ADC and different bandwidths at the signal conditioning stage.}
   \label{fig:noise_fc}
\end{figure}

\begin{figure}
   \centering
   \includegraphics[width=\columnwidth]{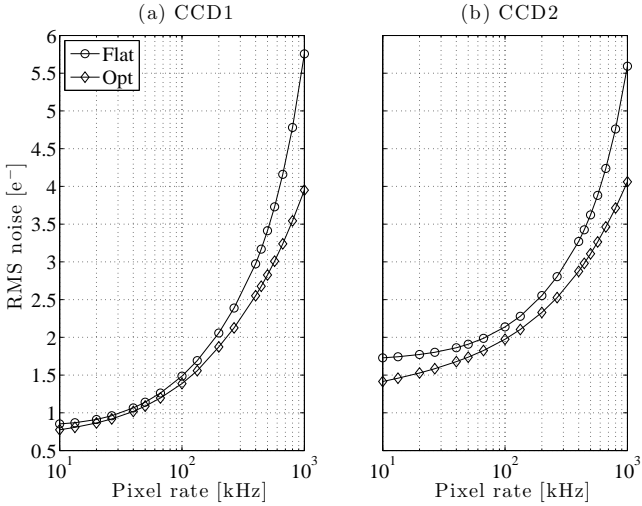} 
   \caption{RMS noise vs pixel rate for both CCDs. The standard averaging filter (Flat) is compared with the optimal filter computed by the proposed method (Opt). The results were generated with a 40 MSPS ADC. For each setup and pixel rate, the bandwidth that produced the lowest noise was selected in order to make a fair comparison of the achievable performance of both methods.}
   \label{fig:comparison}
\end{figure}

Fig. \ref{fig:noises} shows the contribution of all noise sources and the total RMS noise over the pixel rate, taken with a 40 MSPS ADC and a fixed bandwidth for every pixel rate. The theoretical predictions are plotted with solid lines, whereas the error bars were generated with simulations. The simulated results were obtained with the frequency-domain method for noise generation, although the time-domain method produces equivalent results. Each simulation point was computed for 100 pixels and repeated 20 times to compute the mean value and the error bars. The pixel rate is computed as the inverse of the sampling window, so it only depends on the sampling rate and number of samples. The time required for the reset pulse and charge transfer is not considered because it can vary for different CCDs and does not depend on the presented method, so the actual pixel rate is slightly lower. Due to the corner frequency location, white noise is dominant in CCD1 over most of the plotted range, whereas its contribution in CCD2 is dominant below 200 kHz. 

The optimal setup was compared with the standard setup for a DCDS system with flat weights. In the latter, half of the samples are taken at the reset level. After the charge dump, some samples are discarded until the signal is settled to $\Delta/2$ and the remaining samples are used to compute a simple differential average. Fig. \ref{fig:noise_fc} depicts the RMS noise over the pixel rate for both configurations and different bandwidths at the signal conditioning stage. Since the optimal filter is computed as a function of the bandwidth for every pixel rate, the proposed method performs adequately for a typical range of pixel frequencies, even if the bandwidth is fixed. This is an appealing feature, since it does not require to modify electronic components. Furthermore, the proposed method performs better than the averaging filter for any given bandwidth.

The overall optimal setup is reached by selecting the best bandwidth at each pixel rate, which is a result of the semi-exhaustive search depicted in Section \ref{sec:opt}. Fig.~\ref{fig:comparison} shows the RMS noise for both CCDs read out with an averaging filter and with an optimal filter, where the optimal bandwidth was selected independently for both setups in order to make a fair comparison of the achievable performance. The optimal filter noise is always lower, and a significant noise reduction is achieved at high pixel rates due to the use of low bandwidths and the settling period of the CCD. When white and low-frequency noise contributions are commensurable, the
optimal coefficients are successful in lowering the noise floor, particularly at low pixel rates.

\section{Conclusion}
\label{sec:conclusion}

A detailed and thorough mathematical model to describe a DCDS system was presented. Based on this model, the noise statistics at the system output were computed as a function of the CCD parameters and the system design variables. An optimisation model to maximise the SNR was developed, thus providing a systematic design methodology for an optimal DCDS readout system. Theoretical results were compared with realistic simulations to validate the theory and show the potential of the optimisation method. As a result, the trade-offs involved in the design of a DCDS system were analysed and previous experimental disagreements were explained.  

\section*{acknoledgement}
The authors would like to thank Peter Moore and Marco Bonati for their  helpful discussions and suggestions throughout the development of this manuscript. The authors would also like to thank the engineering team at Cerro Tololo International Observatory (CTIO) for providing a Torrent CCD controller and constant technical support.   

This work is funded by the National Commission of Scientific and Technologic Research (CONICYT, Chile) through the FONDECYT project 1130334 and the National Doctoral grant 21140599.
\bibliography{bio}{}
\bibliographystyle{mnras}

\appendix
\section{Pulse shape derivation}
\label{ap:pulse}
An arbitrary two-sided noise power spectrum given by
\begin{eqnarray}
S_\mathrm{m}(i\omega) =A_\mathrm{m}\vert \omega \vert ^{\alpha_\mathrm{m}}
\label{eq:power}
\end{eqnarray}
can be expressed as
\begin{eqnarray}
S_\mathrm{m}(i\omega)&=&A_\mathrm{m}\left( {(i\omega)^{{\alpha_\mathrm{m}}/2}(-i\omega)^{{\alpha_\mathrm{m}}/2}}\right)  \\
           &=& \left(A_\mathrm{m}^{1/2}(i\omega)^{{\alpha_\mathrm{m}}/2}\right)\left(A_\mathrm{m}^{1/2}(i\omega)^{{\alpha_\mathrm{m}}/2}\right)^*.
\label{eq:power2}
\end{eqnarray}
Following the same procedure shown in \cite{pullia2004}, the frequency core pulse is given by 
\begin{eqnarray}
H(i\omega)=A_\mathrm{m}^{1/2}(i\omega)^{{\alpha_\mathrm{m}}/2}
\label{eq:core}
\end{eqnarray}
and the frequency core pulse after a system $G(i\omega)$ can be computed as
\begin{eqnarray}
Y_\mathrm{m}(i\omega)=A_\mathrm{m}^{1/2}(i\omega)^{{\alpha_\mathrm{m}}/2} G(i\omega),
\label{eq:core2}
\end{eqnarray}
which is a hermitian function. The time-domain core pulse can be computed in terms of the system impulse response $g(t)$ and the Fourier derivative property as
\begin{eqnarray}
y_\mathrm{m}(t)=\sqrt{{{A_\mathrm{m}}}}\frac{\mathrm{d}^{{\alpha_\mathrm{m}}/2}}{\mathrm{d}t^{{\alpha_\mathrm{m}}/2}} g(t),
\label{eq:coretime}
\end{eqnarray}
which is a real function. The core pulse is finally scaled in amplitude to make the noise energy consistent with the arrival rate
\begin{eqnarray}
\tilde{y}_\mathrm{m}(t)=\sqrt{{\frac{A_\mathrm{m}}{\nu}}}\frac{d^{{\alpha_\mathrm{m}}/2}}{dt^{{\alpha_\mathrm{m}}/2}} g(t).
\label{eq:coretime}
\end{eqnarray}

\section{Autocorrelation, PSD and stationarity}
\label{ap:auto}

Consider the Fourier transform pair from appendix \ref{ap:pulse}
\begin{eqnarray}
y_\mathrm{m}(t)\rightarrow Y_\mathrm{m}(i\omega).
\end{eqnarray} 
The autocorrelation function of $y_\mathrm{m}(t)$, defined as
\begin{eqnarray}
\mathrm{R}_\mathrm{y}(t_1,t_2)&=& \int_{-\infty}^{\infty}y_\mathrm{m}(\tau-t_1)y_\mathrm{m}(\tau-t_2)d\tau \nonumber\\
				  &=& \int_{-\infty}^{\infty}y_\mathrm{m}(\tau')y_\mathrm{m}(\tau'-(t_2-t_1))d\tau',
\label{eq:corr}
\end{eqnarray}
can be expressed only as a function of $t=t_2-t_1$
\begin{eqnarray}
\mathrm{R}_\mathrm{y}(t)&=& \int_{-\infty}^{\infty}y_\mathrm{m}(\tau)y_\mathrm{m}(\tau-t)d\tau.
\label{eq:corr2}
\end{eqnarray}
If $\mathrm{R}_\mathrm{y}(t)$ is absolutely integrable, its Fourier transform can be computed as
\begin{eqnarray}
S_\mathrm{y}(i\omega)&=&Y_\mathrm{m}(i\omega)Y_\mathrm{m}(i\omega)^* \nonumber\\
		&=&\left(A^{1/2}(i\omega)^{{\alpha}/2}G(i\omega)\right)\left(A^{1/2}(i\omega)^{{\alpha}/2}G(i\omega)\right)^*\nonumber\\
		&=&A\left\vert \omega \right\vert^{\alpha}\left\vert G(i\omega)\right\vert  ^2,
\label{eq:PSD3}
\end{eqnarray} 
which is the noise spectrum of $S_\mathrm{m}(f)$ referred to the ADC input, whereas the full spectrum $S_\mathrm{c}(i\omega)$ can be computed from superposition. Therefore, $S_\mathrm{c}(i\omega)$ is a wide sense stationary (WSS) process if
\begin{eqnarray}
\frac{1}{2\pi} \int_{-\infty}^{\infty}{S_\mathrm{m}}(i\omega)\left\vert G(i\omega)\right\vert^2d\omega < \infty.
\end{eqnarray}
for all $m$. This means that even if $S_\mathrm{m}(i\omega)$ is not WSS, like flicker noise components, the noise at the ADC input can behave as a WSS process if the signal conditioning stage has a highpass filter. Furthermore, even in the absence of a highpass filter, the limited-bandwidth approximation of flicker noise produces the same result.

\label{lastpage}

\end{document}